\documentclass[%
 aip,
 amsmath,amssymb,
 reprint,%
]{revtex4-1}

\usepackage{graphicx}
\usepackage{dcolumn}
\usepackage{bm}
\usepackage[utf8]{inputenc}
\usepackage[T1]{fontenc}
\usepackage{mathptmx}
\usepackage{etoolbox}
\usepackage{xcolor}

\begin{document}
\title{Even-in-magnetic-field part of transverse resistivity as a probe of magnetic transitions}
\author{Antonin Badura}
 \email{badura@fzu.cz}
\affiliation{Faculty of Mathematics and Physics, Charles University, Czech Republic}
\affiliation{Institute of Physics, Czech Academy of Sciences, Czech Republic}
\author{Dominik Kriegner}
\affiliation{Institute of Physics, Czech Academy of Sciences, Czech Republic}
\author{Eva Schmoranzerov{\' a}}
\affiliation{Institute of Physics, Czech Academy of Sciences, Czech Republic}
\author{Karel V{\' y}born{\' y}}
\affiliation{Institute of Physics, Czech Academy of Sciences, Czech Republic}
\author{Miina Leivisk{\" a}}
\affiliation{Institute of Physics, Czech Academy of Sciences, Czech Republic}
\author{Rafael Lopes Seeger}
\author{Vincent Baltz}
\affiliation{Univ. Grenoble Alpes, CNRS, CEA, Grenoble INP, Spintec, F-38000 Grenoble, France}
\author{Daniel Scheffler}
\affiliation{Institute of Physics, Czech Academy of Sciences, Czech Republic}
\author{Sebastian Beckert}
\affiliation{Institut f{\"u}r Festk{\"o}rper- und Materialphysik, Technische Universit{\"a}t Dresden, Germany}
\author{Ismaila Kounta}
\author{Lisa Michez}
\affiliation{Aix-Marseille University, CNRS, CINaM, Marseille, France}
\author{Libor {\v S}mejkal}
\affiliation{Max Planck Institute for the Physics of Complex Systems, Dresden, Germany}
\affiliation{Institut f{\" u}r Physik, Johannes Gutenberg Universit{\"a}t Mainz, Germany}
\affiliation{Institute of Physics, Czech Academy of Sciences, Czech Republic}
\author{Jairo Sinova}
\affiliation{Institut f{\" u}r Physik, Johannes Gutenberg Universit{\"a}t Mainz, Germany}
\affiliation{Institute of Physics, Czech Academy of Sciences, Czech Republic}
\author{Sebastian T. B. Goennenwein}
\affiliation{Universit{\"a}t Konstanz, Fachbereich Physik, 78457 Konstanz, Germany}
\author{Jakub {\v Z}elezn{\' y}}
\affiliation{Institute of Physics, Czech Academy of Sciences, Czech Republic}
\author{Helena Reichlov{\' a}}
\affiliation{Institute of Physics, Czech Academy of Sciences, Czech Republic}

\begin{abstract}
The component of the resistivity tensor $\rho_{ij}$ corresponding to voltage transverse to both an applied current and a magnetic field can be separated into odd and even parts with respect to the applied magnetic field. The former contains information, for example, about the ordinary or anomalous Hall effect. The latter is often ascribed to experimental artefacts and ignored. Here, we show that upon suppressing these artefacts in carefully controlled experiments, useful information remains. We first investigate the well-explored ferromagnet CoFeB, where the even part of $\rho_{yx}$ contains a contribution from the anisotropic magnetoresistance, which we confirm by Stoner--Wohlfarth modelling. We then apply our approach to magnetotransport measurements of $\rm Mn_5Si_3$ thin films, which undergo a transition from non-collinear to an altermagnetic collinear state. In this material, the even part of the transverse signal is sizable only in the low-spin-symmetry phase below $\approx 80$~K. Transverse resistivity measurements thus offer a simple and readily available probe of magnetic order transitions.
\end{abstract}
\maketitle

The measurement of electrical resistivity as a function of an applied external magnetic field is a fundamental characterization tool in solid-state physics. Components of the resistivity tensor $\rho_{ij}$ are typically classified based on their transformation properties under the reversal of the magnetic field $\bm{H}$ or magnetic moments \cite{akgoz1975, Seeman2015}. The antisymmetric components ($\rho^A_{ij} = -\rho^A_{ji}$) arise only when time-reversal symmetry is broken. This can occur either due to an external magnetic field, leading to the ordinary Hall effect, or due to magnetic ordering, resulting in the anomalous Hall effect \cite{nagaosa2010, chang2013, kiyohara2016, smejkal2020}.

The symmetric components are present in all materials and obey the Onsager relations, remaining even under time reversal ($\rho^S_{ij} = \rho^S_{ji}$). Ordinary magnetoresistance (OMR) is a contribution to symmetric $\rho^S_{ij}$ \cite{isasa2016, sailler2024competing}, which leads to a characteristic power-law scaling with an external magnetic field of the resistivity component along the current flow (longitudinal resistivity).  In the transverse resistivity (with respect to the current flow), it is present only if the crystal symmetry is sufficiently low \cite{Seeman2015}. In magnetic materials, the symmetric components also depend on the orientation of the magnetic order vector, an effect known as anisotropic magnetoresistance (AMR). Unlike the anomalous Hall effect, which requires time-reversal symmetry breaking and specific symmetry conditions, AMR is symmetry-allowed in all magnetically ordered materials where the resistivity tensor reflects the orientation of the magnetic order vector  \cite{Ritzinger2022}.

In principle, the various contributions to resistivity can be distinguished based on whether they belong to the symmetric or antisymmetric components of the resistivity tensor. However, the experimental separation of these contributions is not straightforward. Instead, different effects are typically classified based on their odd or even symmetry with respect to the external magnetic field, i.e., as even functions of the field (even-in-field) or odd functions (odd-in-field). Although the anomalous Hall effect and AMR do not intrinsically depend on the applied magnetic field, their experimental detection does since the external field manipulates the magnetic order. This complicates data interpretation, as effects that depend directly on the magnetic field (such as OMR) can mix with those that depend on it indirectly (such as AMR and the AHE). Additionally, effects that depend simultaneously on both the magnetic order and the magnetic field can also exist \cite{Zyuzin2021}.

In typical anomalous Hall effect measurements, the magnetic field is swept perpendicular to the current, and the voltage is measured in a direction orthogonal to both field and current. The AMR is often assumed negligible in such experiments, at least if the magnetization follows the field. However, below the saturation field, the magnetization vector typically deviates from the field direction, invalidating this assumption even in simple ferromagnets. This effect is even more pronounced in low-symmetry materials, such as non-collinear antiferromagnets \cite{fina2014, wang2019, sharma2023planar}, where the magnetization response is more complex. Consequently, the even-in-field component of transverse resistivity can serve as an indicator of AMR. Other even-in-field contributions to transverse resistivity have also been linked to non-centrosymmetric Berry curvature effects \cite{yang2023}. Additionally, a quadratic-in-field term in transverse resistivity can emerge when both the anomalous and ordinary Hall angles are substantial \cite{zhao2023}.

\begin{figure*}[hbt!]
\vspace*{3mm}\includegraphics{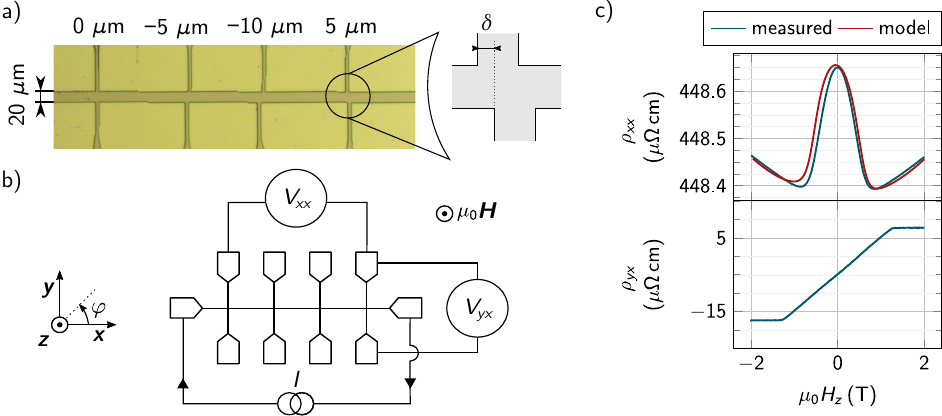}
\caption{Reference CoFeB device: a) A microscope image of transverse contacts with an artificial offset of $\delta$ in the range from $-10\,\rm\mu m$ to $5\,\rm\mu m$, b) schematics of the measurement, c) an example of the dependence of longitudinal $\rho_{xx}$ and transverse $\rho_{yx}$ resistivity on an external magnetic field applied along z- direction at 100 K. The data are shown as measured for $\delta=0\,\rm \mu m$, only the voltage was recalculated to the respective values of resistivity. The panel of longitudinal resistivity also includes the result of the Stoner--Wohlfarth model for the $\rho_{xx}(H_z)$ dependence.}
\label{fig1}
\end{figure*}

From the experimental perspective, artifacts can also contribute to the resistivity measurements. One common source is the misalignment of transverse contacts with respect to the current direction (Fig. \ref{fig1}a), causing an occurrence of the longitudinal signal in the transverse voltage. Since the longitudinal signal is typically dominated by even-in-field contributions, this artifact is also even. To mitigate these artifacts, the even part of the field-sweep data is often subtracted from measured transversal voltage \cite{degrave2013,Reichlova2020}, isolating the dominant Hall signal. However, this also removes potentially valuable information, particularly in materials with complex magnetic structures, where the even-in-field transverse resistivity can reveal details about magnetic order.

Here, we analyze the symmetry of transverse resistivity in thin films of both a simple ferromagnet and a complex magnet which exhibits a transition from non-collinear to an altermagnetic collinear state. By sweeping the out-of-plane magnetic field and measuring both longitudinal and transverse voltages, we identify an even-in-field contribution to transverse resistivity. We show that this contribution persists even with perfectly aligned contacts, does not need to be correlated to the longitudinal voltage, and reflects the underlying magnetic order. 

A ferromagnetic $\rm Co_{40}Fe_{40}B_{20}$ film with a thickness of 15~nm was deposited by magnetron sputtering on a single-crystal MgO (100) substrate using a Bestec UHV deposition system (with a sputtering pressure of $3\cdot 10^{-3}\rm\,mbar$). The film was capped by 2 nm of $\rm Al_2O_3$ deposited via atomic layer deposition to prevent the film's oxidation. The magnetization of CoFeB is approximately $1 200\,\rm kA/m$ at room temperature \cite{Cho2013}. Thin films of $\rm Mn_5Si_3$ were grown by molecular beam epitaxy on intrinsic Si (111) substrates with a $\rm Mn_5Si_3$ thickness ranging from 12 to 20 nm, depending on the sample \cite{Kounta2023}.

Both CoFeB and $\rm Mn_5Si_3$ thin films were patterned into Hall bar microstructures using optical lithography and plasma etching. A detail of a microscope image of the Hall bar prepared on CoFeB is shown in Fig. \ref{fig1}a. This reference device consists of a set of Hall crosses where each pair of transverse contacts has a different artificial offset $\delta$ of $-10\,\rm\mu m$, $-5\,\rm\mu m$, $0\,\rm\mu m$, and $5\,\rm\mu m$. 

A scheme of the experiment geometry is shown in Fig. \ref{fig1}b.  A DC current was applied along the x-axis by a Keithley 2450 source measure unit, and Keithley 2182 nanovoltmeters were used to measure the longitudinal voltage $V_{xx}$ and the transverse voltages $V_{yx}$ on multiple transverse contact pairs. All the presented magnetotransport data were obtained in an Oxford Instruments cryostat Integra AC with a variable-temperature insert and two thermometers to monitor the sample base temperature with high precision. Furthermore, the cryostat was equipped with a superconducting magnet with the sample inserted along the magnet's axis. This allowed the external magnetic field to be swept perpendicular to the sample plane with high precision.

\begin{figure*}[hbt!]
\vspace*{3mm}\includegraphics{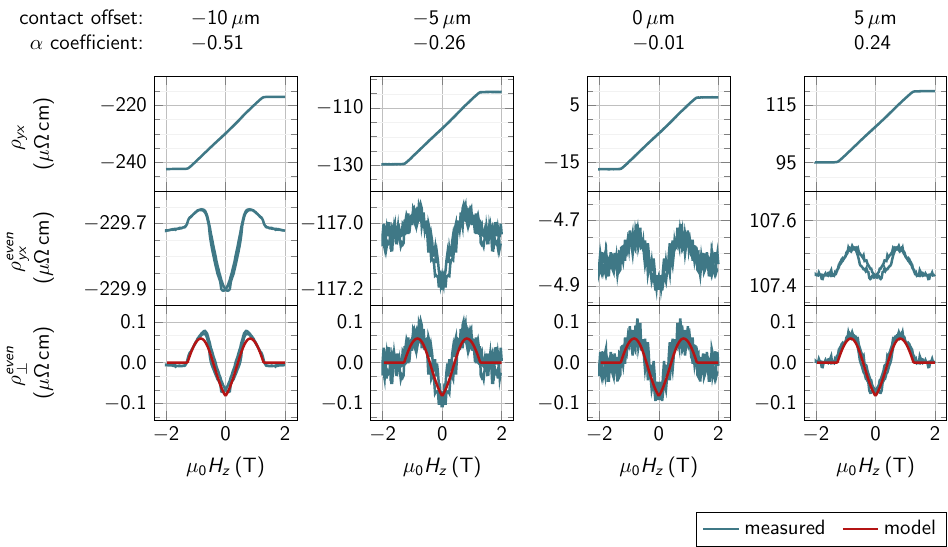}
\caption{Field sweeps measured on a reference CoFeB device at 100 K using multiple contacts with different geometrical offset: The first row shows raw transverse data, the second row its even component, and the third row depicts the offset-independent quantity $\rho_{\perp}^{\rm even}$ with the corresponding $\alpha$ coefficients shown in the header. The last row also shows a model for the even component of $\rho_{yx}$.}
\label{fig2}
\end{figure*}

The crucial step of our proposed method is isolating odd or even components of the measured data with respect to the applied magnetic field. The even and odd components $\rho^{\rm even}$ and $\rho^{\rm odd}$ are defined:
\begin{subequations}
\begin{equation}
\rho_{\leftarrow}^{\rm even}(H) = \frac{\rho_{\leftarrow}(H) + \rho_{\rightarrow}(-H)}{2}
\label{eq_sym-asym-a}
\end{equation}
\begin{equation}
\rho_{\leftarrow}^{\rm odd}(H) = \frac{\rho_{\leftarrow}(H) - \rho_{\rightarrow}(-H)}{2},
\label{eq_sym-asym-b}
\end{equation}
\end{subequations}
where $\rho_\leftarrow(H)$ and $\rho_\rightarrow(H)$ are the subsets of the $\rho$ data corresponding to either descending or ascending sweep direction (i.e. decreasing or increasing magnetic field), which we introduced to account for eventual hysteresis. The corresponding $\rho_{\rightarrow}^{\rm even}(H)$ and $\rho_{\rightarrow}^{\rm odd}(H)$ can be obtained by interchanging $\rho_\leftarrow(H)$ and $\rho_\rightarrow(H)$ in Eq. (\ref{eq_sym-asym-a}) and (\ref{eq_sym-asym-b}). For clarity, the procedure is visualized by Fig. S1 in the supplementary material.\\

An example of longitudinal and transverse resistivity data measured on a 15-nm CoFeB sample is shown in Fig. \ref{fig1}c: The main contributions to the longitudinal resistivity are from AMR and OMR, whereas the transverse resistivity is dominated by the anomalous Hall effect. 

The data measured on a CoFeB Hall bar for various offsets $\delta$ is shown in Fig. \ref{fig2}. The first row corresponds to the measured transverse resistivity $\rho_{yx}(H_z)$: The dominant part of the field dependence is the odd component corresponding to the anomalous Hall effect (AHE) measured along the magnetic hard axis and, therefore, showing no hysteresis. The vertical offset of the data reflects the geometrical misalignment. There is also a clear even part, as seen in the second line, which shows the even component of transverse resistivity $\rho_{yx}^{\rm even}(H_z)$. The even component depends on the geometrical offset as expected since it contains a contribution from longitudinal resistivity dominated by OMR and AMR. We, therefore, introduce a quantity that is independent of the geometrical misalignment $\rho_{\perp}^{\rm even}$:
\begin{equation}
\rho_{\perp}^{\rm even}(H) = \rho_{yx}^{\rm even}(H) - \alpha\cdot\rho_{xx}^{\rm even}(H).
\label{eq_rho-perp}
\end{equation}
In this definition, the quantity $\alpha={\langle\rho_{yx}^{\rm even}\rangle}/{\langle\rho_{xx}^{\rm even}\rangle}$ is the ratio of the $\rho_{yx}^{\rm even}$ and $\rho_{xx}^{\rm even}$ mean values with respect to the field $H$. $\alpha$ quantifies the projection of longitudinal resistivity in the transverse resistivity. By subtracting the even component of longitudinal resistivity $\rho_{xx}^{\rm even}$ scaled by $\alpha$, only the even component of transverse resistivity free of any geometrical misalignment and constant offset is left. Please note that the arithmetic mean is only one of many possibilities of quantifying the $\rho_{yx}^{\rm even}$ and $\rho_{xx}^{\rm even}$ in the definition of the $\alpha$ coefficient. However, the choice does not influece the resulting $\rho_{\perp}$ in our case, as the relative variations of $\rho_{yx}^{\rm even}(H)$ and $\rho_{xx}^{\rm even}(H)$ are small.

The last row of Fig. \ref{fig2} reveals that $\rho_{\perp}^{\rm even}$ is identical for all transverse contacts as expected and does not depend on the geometrical offset. Because in our polycrystalline sample, the contribution of OMR is not expected in transverse resistivity, the origin of the remaining signal can be attributed to AMR. This is also in agreement with the saturation of $\rho_\perp^{\rm even}$ at high magnetic fields. Vice versa, the presence of a finite transverse AMR in our data reveals that the magnetization does not follow the external magnetic field entirely. The field dependence of the longitudinal AMR and the transverse AMR in our polycrystalline sample is distinct (compare Fig. 1c and the last row of Fig. 2) because the two effects can exhibit a different dependence on the magnetization orienation \cite{limmer2008advanced}.

\begin{figure*}[hbt!]
\vspace*{1mm}\includegraphics{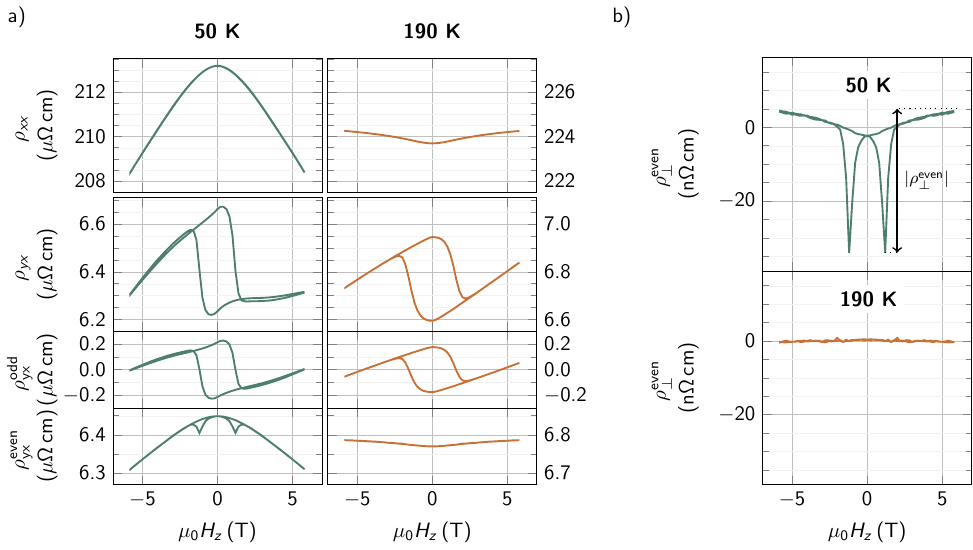}
\caption{Field sweeps measured on $\rm Mn_5Si_3$ epitaxial films at 50~K and 190~K: a) Measured longitudinal and transverse resistivity $\rho_{xx}$ and $\rho_{yx}$ for two temperatures. Odd and even components of $\rho_{yx}$ with respect to the applied magnetic field are separated. b) Misalignment-independent quantity $\rho_{\perp}^{\rm even}$.}
\label{fig3}
\end{figure*}

In order to describe the origin of $\rho_{\perp}^{\rm even}$ in CoFeB, we use a single domain model to describe the orientation of the magnetization vector as a function of field strength. We obtain the uniaxial anisotropy constant $K$ by fitting the dependence of longitudinal resistivity on the applied magnetic field direction within the Stoner--Wohlfarth model. We allow the magnetization to develop a finite projection in the sample plane. The magnetization vector orientation is described by the angles $\theta$ and $\varphi$: $\theta$ describes the angle between the $z$-axis and the magnetization, and the in-plane direction is determined by $\varphi$ as defined in Fig.~1b. The magnetization is aligned with the out-of-plane external magnetic field at high magnetic fields. Once the magnetic field is reduced below the anisotropy field (1.6~T, see Fig.~1c), the magnetization vector continuously cants towards the sample plane, and it concomitantly changes its in-plane orientation constrained by the weak in-plane anisotropy. In this notation, we calculate transverse resistivity $\rho_{yx}$ as follows:
\begin{equation}
\rho_{yx}(\theta, \varphi) = a_1 \cos\theta + a_2\cos^3\theta + a_3\sin^2\theta\sin\varphi\cos\varphi,
\label{eq_rhoxy}
\end{equation}
where $a_{1,2,3}$ are coefficients which we identified with the following values using the Stoner--Wohlfarth model: $a_1 = 11.62\,\rm\mu\Omega cm$, $a_2 = -0.28\,\rm\mu\Omega cm$, $a_3 = 1.00\,\rm\mu\Omega cm$. The first two terms in Eq. (\ref{eq_rhoxy}) describe the anomalous Hall effect and its anisotropy \cite{Zhang2011}, whereas the $a_3$ coefficient quantifies the contribution of transverse AMR \cite{Ritzinger2022}. The value of $a_3$  we found falls within the range typical for CoFeB alloys (for example, $\rm Co_{60}Fe_{20}B_{20}$ was reported to have $2.5\rm\,\mu\Omega cm$  \cite{Seemann2011}). The finite magnetization projection in the sample plane is required to understand the field dependence of $\rho_{\perp}^{\rm even}$, as shown in Fig. \ref{fig2} (the red line is the model). Although the longitudinal and the transverse signal shapes are seemingly very different, the same magnetization trajectory can also fit the longitudinal AMR (see Fig. \ref{fig1}c). Both the measured data and the modelling confirm that the AMR is non-zero, although the magnetic field is perpendicular to the current and voltage detection direction.
\\

The AMR in ferromagnets can be measured and identified because it saturates with the saturation of the magnetization. The AMR is, however, not so well understood in materials with complex spin texture, and it is challenging to distinguish the AMR from the OMR because antiferromagnets or altermagnets do not necessarily saturate in an achievable magnetic field. If we apply current along a high-symmetry crystallographic axis, the OMR does not contribute to $\rho_\perp^{\rm even}$ \cite{Seeman2015}. The quantity $\rho_\perp^{\rm even}$ could, therefore, serve as a good indicator of more complex magnetoresistance signals beyond OMR. In the following, we test this approach in altermagnetic $\rm Mn_5Si_3$, which exhibits magnetic phase transitions \cite{Reichlova2020, Surgers2014, leiviska2024anisotropy}. Our $\rm Mn_5Si_3$ films show a transition between a low-temperature non-collinear magnetic phase AM1 and a high-temperature collinear magnetic phase AM2 at 70~K. The films become paramagnetic at 240~K \cite{Reichlova2020}. It was shown that due to their particular spin and crystal symmetry, the epitaxial $\rm Mn_5Si_3$ thin films fulfill the altermagnetic symmetry requirements \cite{Reichlova2020, Smejkal2022} and can exhibit anomalous Hall effect despite their vanishing magnetization \cite{Reichlova2020}. Interestingly, the Hall effect (the odd part of the transverse voltage) has the same magnitude in the whole temperature range (10--240~K) and, therefore, is insensitive to the magnetic phase transitions \cite{Reichlova2020}. 

\begin{figure*}[hbt!]
\vspace*{3mm}\includegraphics{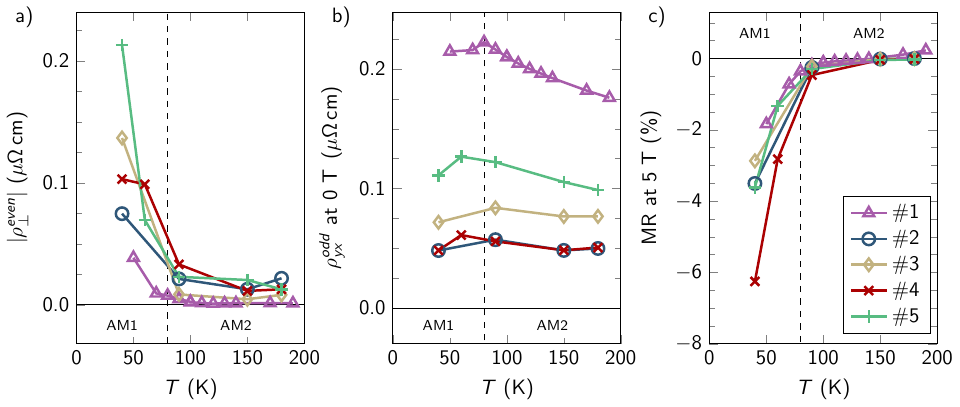}
\caption{Temperature dependence of a) the maximum change of the misalignment-independent quantity $\rho_{\perp}^{\rm even}(H)$ as defined in Fig. \ref{fig3}b, b) the magnitude of spontaneous Hall resistivity, i.e. the value of $\rho_{yx}^{\rm odd}$ in zero magnetic field,  and c) magnetoresistance $(\rho_{xx}(5{\rm\,T})-\rho_{xx}(0{\rm\,T}))/\rho_{xx}(0{\rm\,T})$ for five $\rm Mn_5Si_3$ samples. The data in Fig.~3a are from the sample \#1.}
\label{fig4}
\end{figure*}

An example of transverse resistivity $\rho_{yx}(H_z)$ measured at 190~K and 50~K in a $\rm Mn_5Si_3$ thin film is shown in Fig.~\ref{fig3}a, together with its odd and even components with respect to the magnetic field. For illustration, also the longitudinal resistivity $\rho_{xx}(H_z)$ is included in Fig.~\ref{fig3}a. It can be seen that the transverse resistivity has a clear even component pronounced in the low temperature (low spin symmetry) regime. To remove the effect of geometrical misalignment, we use the approach described above, and we evaluate the offset-independent value $\rho_{\perp}^{\rm even}$. This component is substantially higher at 50~K than at 190~K, as shown in Fig.~\ref{fig3}b. The hysteresis behaviour observed in the $\rho_{\perp}^{\rm even}$ highlights its difference to the $\rho_{xx}$. In the hexagonal lattice of $\rm Mn_5Si_3$, in the absence of the magnetic order, and for current along a high symmetry direction, the transverse OMR is not allowed by symmetry \cite{Gallego2019}. Therefore the $\rho_{\perp}^{\rm even}$  must arise from the magnetic properties of the material. We note that $\rho_{\perp}^{\rm even}$ cannot result from a potential misalignment of the sample in the cryostat, which would cause a finite in-plane projection of the external magnetic field. Such a misalignment would need to be unrealistically large to induce a sizable $\rho_{\perp}^{\rm even}$ compared to $\rho_{yx}^{\rm odd}$ (see Fig.~\ref{fig3}, 50~K).

The idea to use $\rho_{\perp}^{\rm even}$ as a tracer for magnetic properties reveals its potential if we apply it to a set of temperatures and samples. We evaluated the $\rho_{\perp}^{\rm even}$ in the temperature range of 10--190~K for several $\rm Mn_5Si_3$ samples that differ in their parameters, such as layer thickness or composition of spurious phases \cite{Reichlova2020}. In Fig. \ref{fig4}a, we show the absolute amplitude of the $\rho_{\perp}^{\rm even}(H_z)$ dependence extracted directly from the detected hysteresis loops (see Fig. \ref{fig3}b) for different sample temperatures. The signal is sizable in the AM1 phase. Unlike the usual treatment of the transverse resistivity, i.e. considering the spontaneous or saturated Hall signal $\rho_{yx}^{\rm odd}$ (see Fig. \ref{fig4}b), our even signal $\rho_{\perp}^{\rm even}$ reflects the phase transition precisely, and is very small in the AM2 phase (above 70~K). 

The presence of $\rho_{\perp}^{\rm even}$ in the AF1 phase may result from the sizable transverse AMR, attributed to the low spin symmetry of this phase. Furthermore, the presence of even-in-field AHE contributions to $\rho_{\perp}^{\rm even}$ below the saturation field cannot be ruled out. The presence of magnetoresistance related to the magnetic order in $\rho_{\perp}^{\rm even}$ would be, in principle, in agreement with the longitudinal resistivity $\rho_{xx}$ signals.  Interestingly, when comparing multiple samples, $\rho_{xx}$ and $\rho_{\perp}^{\rm even}$ cannot be correlated as follows from Fig. \ref{fig4}: The maximal value of the longitudinal magnetoresistance was observed in a different sample than the maximum of $\rho_{\perp}^{\rm even}$. We would like to point out that the method is limited only to magnetic magnetic where the magnetic phase transition is accompanied by the corresponding change of their magnetotransport response.

In the Hall geometry, the even part of transverse resistivity does not necessarily result from measurement artefacts, such as geometrical misalignment. Rather, anisotropic magnetoresistance can yield a non-trivial even-in-field transverse resistivity response. To extract these, we define the even part of the transverse resistivity $\rho_{\perp}^{\rm even}$ which is independent of the geometrical offset, and show that $\rho_{\perp}^{\rm even}$ can contain useful information about the magnitude of the AMR. We validate this approach in a simple polycrystalline ferromagnet, where we approximate the $\rho_{\perp}^{\rm even}$ by a single domain model. We show that longitudinal and transverse AMR measured in the magnetic field sweep do not have to show the same magnitude and symmetry due to the complex magnetization trajectory. We further then apply our approach to $\rm Mn_5Si_3$, which transitions from non-collinear to an altermagnetic collinear state. We isolate $\rho_{\perp}^{\rm even}$ signals in a series of measurements in $\rm Mn_5Si_3$ thin film samples, and we show that the $\rho_{\perp}^{\rm even}$ can serve as a probe of the magnetic phase transition in the material. The different magnitude of $\rho_{\perp}^{\rm even}$ in low- and high-temperature magnetic phases of $\rm Mn_5Si_3$ is in agreement with the proposed magnetic ordering of the thin films \cite{Reichlova2020}.

\section*{Supplementary material}
The supplementary material contains a detailed description of the procedure for separating the odd-in-field and even-in-field components of resistivity signals. 

\begin{acknowledgments}
We acknowledge CzechNanoLab Research Infrastructure, supported by MEYS CR (LM 2018110) and LNSM-LNSpin and the French national research agency (ANR), grant No. ANR-20-CE92-0049-01. Financial support was provided by the Deutsche Forschungsgemeinschaft (DFG, German Research Foundation) via Project-ID 445976410 – GO 944/8, 490730630 GO 944/10, 425217212 – SFB 1432, and 45976410. The study was supported by Charles University, project GA UK No. 266723. HR was supported by the Grant Agency of the Czech Republic Grant No. 22-17899K, TERAFIT - CZ.02.01.01/00/22$\_$008/0004594 and the Dioscuri Program LV23025 funded by MPG and MEYS. D.K. acknowledges the support from the Czech Academy of Sciences (project No. LQ100102201).
\end{acknowledgments}

\section*{Data Availability Statement}
The data that support the findings of this study are available from the corresponding author upon reasonable request.

\bibliography{refsymVxy.bib}
\end{document}